\documentclass[12pt]{article}

\usepackage{latexsym}
\usepackage{graphicx}
\usepackage{rotating}
\usepackage{indentfirst}

\makeatletter
\renewcommand\section{\@startsection {section}{1}{\z@}%
                                   {-3.5ex \@plus -1ex \@minus -.2ex}%
                                   {2.3ex \@plus.2ex}%
                                   {\normalfont\small\bfseries}}
\renewcommand\subsection{\@startsection{subsection}{2}{\z@}%
                                   {3.25ex \@plus 1ex \@minus .2ex}
                                   {1.5ex \@plus.2ex}%
                                   {\normalfont\small\bfseries}}

\makeatother

\def\Rmath{\mbox{\hbox{ I\hskip -2.pt R}}}
\def\Cmath{\mbox{\hbox{ l\hskip -5.8pt C\/}}}
\def\emath{\mbox{e}}

\begin{document}


\begin{center}

\begin{figure}
\vspace{-1.6cm}
\includegraphics[width=15.9cm,height=4.3cm]{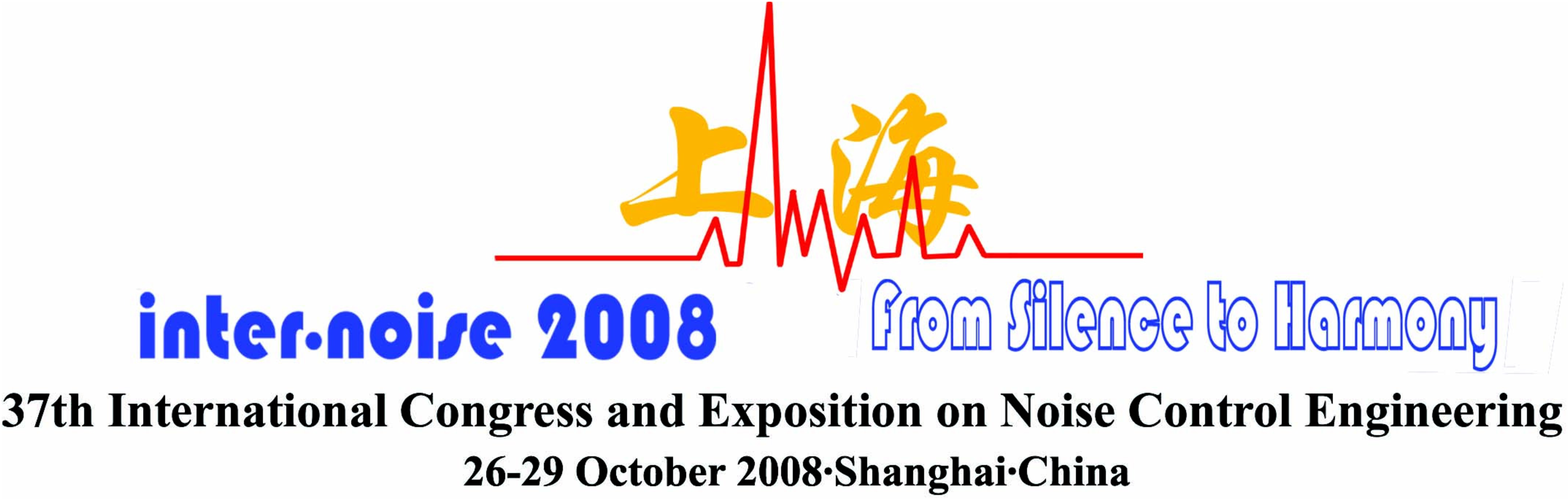}
\end{figure}

 {\bf \large Multiple Resonances in Fluid-Loaded Vibrating Structures }  ~\vspace{2ex}~ \\
{\large P.-O.~Mattei\footnote{Email address: mattei@lma.cnrs-mrs.fr}}  \\
{ Laboratoire de M\'ecanique et d'Acoustique} \\
{ 31 chemin Joseph Aiguier, 13402 Marseille Cedex 20, France } ~\vspace{3ex}~ \\

\noindent{\bf ABSTRACT}\\
~\vspace{1ex}~ 
\begin{minipage}[t]{12.5cm}
{\footnotesize
This study deals with vibroacoustics under heavy fluid loading conditions. Considerable attention has been and remains focused on this subject not only because industry is very concerned but also because of mathematical difficulties that make the numerical resolution of the problem very difficult.

It was recently observed~\cite{POM-2007} in a numerical study on a high order perturbation method under heavy fluid loading that a loaded vibrating plate results in a frequency shift of the in vacuo single resonance (in both the  real part because of the fluid added mass and the imaginary part because of energy lost by radiation into the fluid) as well as increase in the number of the resonance frequencies : as a result of the loading, each single in vacuo resonance frequency of the structure is transformed into a multiple resonance frequency. Here we show that this phenomenon is said to be an extension to the heavy loading condition of the Sanchez's classical result that have established that in the case of a light loading conditions ``the scattering frequencies of a fluid loaded elastic structure (ie the resonance frequencies) are nearly the real eigenfrequencies of the elastic body alone and the complex scattering frequencies of the fluid with a rigid solid''.
          
Using classical results in the framework of the theory of entire functions, it is established that a single resonance of a simply supported fluid loaded rectangular plate is transformed into an infinite number of resonances.}
\end{minipage}
\end{center}


\section{INTRODUCTION}

\indent It is well known that the need to accurately describe fluid structure interactions increases the computational cost of the various numerical methods used. Most numerical methods can be used to deal with the resolution of linear systems of simultaneous equations. When the loading conditions are taken into account, the size of the matrices involved increases considerably, and they become full and frequency dependent. Any method of reducing these drawbacks is therefore most welcome; and one of the best methods available is that based on the perturbation approach. Asymptotic analysis can be conduced when the loading is light, as in the case of a metal structure surrounded by air. This approach consists in introducing a small parameter $\epsilon$, which is the ratio between the surface mass density and the fluid density. Using perturbation expansions~\cite{Nayfeh}, one can then construct an approximate solution, based on the {\it in vacuo} eigenmodes, which is not only easier to calculate than the solution of the exact problem but also leads to a better understanding of the various phenomena involved~\cite{PJTF-DH-POM-CM}. 

The light fluid loading perturbation method (involving high density structures in contact with a small density fluid such as aluminum plate in contact with air) was recently extended~\cite{POM-2007} to cases where the perturbation parameter becomes large (which corresponds to a very light and/or very thin structure in contact with a high density fluid). During this study multiple resonance processes were observed where a single in vacuo resonance mode can have several resonance frequencies under heavy fluid loading. The spectrum of the operator can no longer be described in terms of the discrete spectrum based on a countable series of resonance mode/resonance frequency pairs. This typically nonlinear phenomenon is closely linked to the concept of non-linear frequency modes~\cite{Dazel-Lamarque-Sgard}, where the non linearity depends on the time parameters (such as those involved in porous materials) rather than on the more usual geometrical parameters. This frequency non-linearity is related to problems which are written in the form ${\cal O} (\omega) U = S$, where $u$ is the unknown factor and $S$ is the source term.  ${\cal O} (\omega)$ is an operator which depends non linearly on the angular frequency $\omega$, as in the case of porous environments or in the context of vibroacoustics where the coupling depends non-linearly on the frequency via the Helmholtz equation Green's kernel. When the coupling is weak (as in air), this process is not easily observed.

Here it is established that this behavior is a general property of fluid-loaded plates. In paragraph~\ref{perturbation}, the approximate method for calculating the spectrum of a fluid loaded vibrating structure based on a high order perturbation expansion is recalled, and this multiple resonance is property is then illustrated by giving a numerical example which confirms the validity of the method (based on comparisons with a numerical resolution of the exact equations). Paragraph~\ref{multiple} gives a theoretical description of multiple resonance in the case of the simply supported plate loaded with a fluid with an arbitrarily high density. It is then established, using classical results on the connection between the order of an entire function and the distribution of its zeros, that any in vacuum resonance mode in this structure will have an infinite number of resonances under heavy fluid loading. The conclusions and the possible extensions of this work are presented in paragraph~\ref{conclusions}. 

~

\section{HIGH ORDER PERTURBATION METHOD} \label{perturbation}

Let us first recall the main result obtained in our previous study~\cite{POM-2007}. Consider a thin finite elastic structure (with surface density $\rho_p h$ and thickness $h$) occupying a domain $\Sigma$, the behavior of which is described by a differential operator ${\cal A}$. This structure is loaded with a perfect fluid at rest extending to infinity. Using perturbation methods, one calculates the eigenmodes $\tilde{U}_m(M,\omega)$, where $M$ is a point of the domain $\Sigma$, and eigenvalues $\tilde{\Lambda}_m(\omega) =\rho_p h \tilde{\omega}_m^2 (\omega)$ as solutions of the classical integrodifferential equation
\begin{equation}
{\cal A} \tilde{U}_m(M,\omega) - \tilde{\Lambda}_m(\omega) \left( \tilde{U}_m(M,\omega) - \epsilon \int_{\Sigma}  \tilde{U}_m(M',\omega) G(M;M',\omega) dM' \right) = 0 \label{weak_u}
\end{equation}
where $\epsilon = \rho_f/\rho_p h$ is a small parameter in the case of a metal structure in contact with air (density $\rho_f$). $G(M;M',\omega)$ is the Green's function of the Neumann problem in the Helmholtz equation outside the surface $\Sigma$. One defines the radiation impedance $\beta_{\omega} (U,V) = \int_{\Sigma} \int_{\Sigma}  U(M') G(M;M',\omega) V^*(M) d M d M'$, where $U$ and $V$ are function defined on the surface $\Sigma$. The weak formulation of the problem then reads find $\tilde{U}_m$ such that for every $V$ $a(\tilde{U}_m,V^*) - \tilde{\Lambda}_m(\omega) \left( \langle \tilde{U}_m,V^* \rangle - \epsilon  \beta_{\omega} (\tilde{U}_m,V^*)  \right) =  0$, where $a(U,V^*)$ stands for the potential elastic energy of the structure, $\langle U,V^* \rangle$ is the usual inner product, $\rho_p h \omega^2 \langle U,V^* \rangle$ is the kinetic energy and $\rho_p h \omega^2 \beta_{\omega} (\tilde{U}_m,V^*)$ is the energy lost by radiation into the fluid. It is worth noting that because of the fluid-loading, each eigenmode and each eigenvalue depends on the frequency. Until necessary, this dependence is omitted to facilitate reading.

Let us expand into a Rayleigh-Schr\"odinger perturbation series~\cite{Nayfeh} both the eigenmodes $\tilde{U}_m$ and the eigenvalues $\tilde{\Lambda}_m$ in $\epsilon$ : 
\begin{eqnarray}
\tilde{U}_m (M) & = & \tilde{U}_m^{(0)}(M) + \epsilon \tilde{U}_m^{(1)} (M) + \cdots + \epsilon^s  \tilde{U}_m^{(s)}(M) + \cdots \\
\tilde{\Lambda}_m & = & \tilde{\Lambda}_m^{(0)} + \epsilon \tilde{\Lambda}_m^{(1)} + \cdots + \epsilon^s \tilde{\Lambda}_m^{(s)} +  \cdots
\end{eqnarray}
The Rayleigh-Schr\"{o}dinger Method is one of the most powerful methods available to dealing with vibroacoustics problems involving low frequency dependent damping such as those arising in viscosity or sound radiation in light fluid. Substituting the two perturbation expansions into the weak formulation given by equation~(\ref{weak_u}), collecting and equating the coefficients of equal power $\epsilon$ to zero yields:
\begin{eqnarray}
\epsilon^0 : a(\tilde{U}_m^{(0)},V) - \tilde{\Lambda}_m^{(0)}  \langle \tilde{U}_m^{(0)},V \rangle & = & 0, \label{e_pf_0} \\
\epsilon^0 : a(\tilde{U}_m^{(1)},V) - \tilde{\Lambda}_m^{(0)}  \langle \tilde{U}_m^{(1)},V \rangle & = & \tilde{\Lambda}_m^{(1)}  \langle \tilde{U}_m^{(0)},V \rangle - \tilde{\Lambda}_m^{(0)} \beta_{\omega} ( \tilde{U}_m^{(0)},V ), \label{e_pf_1} \\
 & \vdots & \nonumber \\
\epsilon^s : a(\tilde{U}_m^{(s)},V) - \tilde{\Lambda}_m^{(0)}  \langle \tilde{U}_m^{(s)},V \rangle & = &  \sum_{l=1}^{l=s} \tilde{\Lambda}_m^{(l)}  \langle \tilde{U}_m^{(s-l)},V \rangle - \sum_{l=0}^{l=s-1} \tilde{\Lambda}_m^{(l)} \beta_{\omega} ( \tilde{U}_m^{(s-l-1)},V ),  \label{e_pf_s}
\end{eqnarray}

Equation~(\ref{e_pf_0}) is the usual relation giving the eigenmodes of the elastic structure {\it in vacuo}. If the eigenmodes $\tilde{U}_m^{(0)}(M)$ of the elastic plate {\it in vacuo} are normalized, it is easy to obtain the first order term of the eigenvalue form equation~(\ref{e_pf_1}), that is
\begin{equation}
\tilde{\Lambda}_m^{(1)} = \tilde{\Lambda}_m^{(0)}  \beta_{\omega }^{mm},
\end{equation}
in the previous equation, it has been denoted $\beta_{\omega }^{mn} = \beta_{\omega }( \tilde{U}_m^{(0)}, \tilde{U}_n^{(0)*} )$. To simplify the reading this notation will be used in the rest of the paper. $\tilde{U}_m^{(1)} (M)$ is calculated as an expansion into a series of zeroth-order modes: $\tilde{U}_m^{(1)}(M) = \sum_n \alpha_m^{1n} \tilde{U}_n^{(0)} (M)$ where the coefficients $\alpha_m^{1n}$ are easy to calculate~\cite{Nayfeh}. One has
\begin{equation}
\alpha_m^{1n} = \frac{ \tilde{\Lambda}_m^{(0)} }{\tilde{\Lambda}_m^{(0)} - \tilde{\Lambda}_n^{(0)}} \beta_{\omega }^{mn}.
\end{equation}

For all $s>1$, one has:
\begin{equation}
\tilde{\Lambda}_m^{(s)} = \sum_{l=0}^{l=s-1} \tilde{\Lambda}_m^{(l)} \beta_{\omega} ( \tilde{U}_m^{(s-l-1)},\tilde{U}_m^{(0)} ) - \sum_{l=1}^{l=s-1} \tilde{\Lambda}_m^{(l)} \langle \tilde{U}_m^{(s-l)},\tilde{U}_m^{(0)} \rangle.
\end{equation}
In a similar manner, each component of the eigenmode's perturbation expansion is developed into a series of zeroth-order modes:
\begin{equation}
\tilde{U}_m^{(s)}(M) = \sum_n \alpha_m^{sn} \tilde{U}_n^{(0)} (M).
\end{equation}

One shows that the expansions of the eigenvalues are given by:
\begin{eqnarray}
\frac{ \tilde{\Lambda}_m^{(2)} }{ \tilde{\Lambda}_m^{(0)} } & = & (\beta_{\omega }^{mm})^2 + \sum_{p \neq m} \alpha_m^{1p}\beta_{\omega }^{mp}, \\
\frac{ \tilde{\Lambda}_m^{(3)} }{ \tilde{\Lambda}_m^{(0)} } & = & (\beta_{\omega }^{mm})^3 + 2 \beta_{\omega }^{mm} \sum_{p \neq m} \alpha_m^{1p} \beta_{\omega }^{mp}  + \sum_{p \neq m} \alpha_m^{2p} \beta_{\omega }^{mp}, \\
\frac{ \tilde{\Lambda}_m^{(4)} }{ \tilde{\Lambda}_m^{(0)} } & = & (\beta_{\omega }^{mm})^4 + 3 (\beta_{\omega }^{mm})^2 \sum_{p \neq m} \alpha_m^{1p} \beta_{\omega }^{mp}  + 2 \beta_{\omega }^{mm} \sum_{p \neq m} \alpha_m^{2p} \beta_{\omega }^{mp} + \sum_{p \neq m} \alpha_m^{3p} \beta_{\omega }^{mp} \nonumber \\
& & + \left(\sum_{p \neq m} \alpha_m^{1p} \beta_{\omega }^{mp} \right)^2 - \alpha_m^{2m} \sum_{p \neq m} \alpha_m^{1p} \beta_{\omega }^{mp}, \\
\frac{ \tilde{\Lambda}_m^{(5)} }{ \tilde{\Lambda}_m^{(0)} } & = & (\beta_{\omega }^{mm})^5 + 4 (\beta_{\omega }^{mm})^3 \sum_{p \neq m} \alpha_m^{1p} \beta_{\omega }^{mp}  + 3 ( \beta_{\omega }^{mm})^2 \sum_{p \neq m} \alpha_m^{2p} \beta_{\omega }^{mp} + 2 \beta_{\omega }^{mm} \sum_{p \neq m} \alpha_m^{3p} \beta_{\omega }^{mp} \nonumber \\
& & + 3 \beta_{\omega }^{mm} \left(\sum_{p \neq m} \alpha_m^{1p} \beta_{\omega }^{mp} \right)^2 - 2 \beta_{\omega }^{mm} \alpha_m^{2m} \sum_{p \neq m} \alpha_m^{1p} \beta_{\omega }^{mp} \nonumber \\
& & + \sum_{p \neq m} \alpha_m^{4p} \beta_{\omega }^{mp} + 2 \sum_{p \neq m} \alpha_m^{1p} \beta_{\omega }^{mp} \sum_{p \neq m} \alpha_m^{2p} \beta_{\omega }^{mp} \nonumber \\
& & - \sum_{p \neq m} \alpha_m^{1p} \alpha_m^{3p} \beta_{\omega }^{mp}- \alpha_m^{2m} \sum_{p \neq m} \alpha_m^{2p} \beta_{\omega }^{mp}.
\end{eqnarray}

Now let us calculate the eigenvalue expansion up to order 5:
\begin{equation}
\frac{\tilde{\Lambda}_m }{ \tilde{\Lambda}_m^{(0)}} \approx 1 + \sum_{s=1}^{s=5} \epsilon^s \frac{ \tilde{\Lambda}_m^{(s)} }{ \tilde{\Lambda}_m^{(0)}} + \cdots. \label{sum_order_5}
\end{equation}
Upon re-ordering the terms in equation~(\ref{sum_order_5}), one obtains:
\begin{eqnarray}
\frac{ \tilde{\Lambda}_m }{ \tilde{\Lambda}_m^{(0)} } & \approx & \left[ 1 + \epsilon \beta_{\omega }^{mm} + \epsilon^2 (\beta_{\omega }^{mm})^2 + \epsilon^3 (\beta_{\omega }^{mm})^3 + \epsilon^4 (\beta_{\omega }^{mm})^4 + \epsilon^5 (\beta_{\omega }^{mm})^5 + \cdots \right] \nonumber \\
& & + \epsilon^2 \sum_{p \neq m} \alpha_m^{1p} \beta_{\omega }^{mp} \left[ 1 + 2 \epsilon \beta_{\omega }^{mm} + 3 \epsilon^2 (\beta_{\omega }^{mm})^2 + 4 \epsilon^3 (\beta_{\omega }^{mm})^3 + \cdots \right] \nonumber \\
& & + \epsilon^3 \sum_{p \neq m} \alpha_m^{2p} \beta_{\omega }^{mp} \left[ 1 + 2 \epsilon \beta_{\omega }^{mm} + 3 \epsilon^2 (\beta_{\omega }^{mm})^2 + \cdots \right] \nonumber \\
& & + \epsilon^4 \left\{ \sum_{p \neq m} \alpha_m^{3p} \beta_{\omega }^{mp} \left[ 1 + 2 \epsilon \beta_{\omega }^{mm} + \cdots \right] + \left(\sum_{p \neq m} \alpha_m^{1p} \beta_{\omega }^{mp} \right)^2 \left[ 1 + 3 \epsilon \beta_{\omega }^{mm} + \cdots \right] \right. \nonumber \\
& & \hspace{3em} \left.  - \alpha_m^{2m} \sum_{p \neq m} \alpha_m^{1p} \beta_{\omega }^{mp} \left[ 1 + 2 \epsilon \beta_{\omega }^{mm} + \cdots \right] \right\} \nonumber \\
& & + \epsilon^5 \cdots  \label{order_5_original}
\end{eqnarray}

For $\epsilon \beta_{\omega }^{mm} <1$, one can recognize in each square bracket in equation~(\ref{order_5_original}) the expansion of $1/(1 -\epsilon \beta_{\omega }^{mm})$ and its successive powers. By identifying the expansion with the original functions, one obtains :

\begin{eqnarray}
\frac{ \tilde{\Lambda}_m }{ \tilde{\Lambda}_m^{(0)} } & \approx & \frac{1}{(1 - \epsilon \beta_{\omega }^{mm})} + \epsilon^2 \frac{\sum_{p \neq m} \alpha_m^{1p} \beta_{\omega }^{mp}}{(1 - \epsilon \beta_{\omega }^{mm} )^2} + \epsilon^3  \frac{\sum_{p \neq m} \alpha_m^{2p} \beta_{\omega }^{mp}}{(1 - \epsilon \beta_{\omega }^{mm} )^2} \nonumber \\
& & +\epsilon^4 \left\{  \frac{\sum_{p \neq m} \alpha_m^{3p} \beta_{\omega }^{mp}}{(1 - \epsilon \beta_{\omega }^{mm} )^2} +  \frac{\left(\sum_{p \neq m} \alpha_m^{1p} \beta_{\omega }^{mp} \right)^2}{(1 - \epsilon \beta_{\omega }^{mm} )^3} - \alpha_m^{2m}  \frac{\sum_{p \neq m} \alpha_m^{1p} \beta_{\omega }^{mp}}{(1 - \epsilon \beta_{\omega }^{mm} )^2} \right\} \nonumber \\
& & + \epsilon^5 \cdots  \label{order_5_modified}
\end{eqnarray}

Now by limiting this expansion to the first term in equation~(\ref{order_5_modified}), one obtains the high order perturbation expansion:

\begin{equation}
\frac{ \tilde{\Lambda}_m }{ \tilde{\Lambda}_m^{(0)} } \approx  \frac{1}{(1 - \epsilon \beta_{\omega }^{mm})}.\label{high_order_resummation}
\end{equation}
It is worth noting that while written as a function of $\epsilon$ alone, the relation (\ref{high_order_resummation}) can not be considered as a first order expansion since it takes into account an infinite expansion into powers of $\epsilon$.

Now by introducing the pulsations, one obtains the expression of the high order, frequency dependent, eigenpulsations:
\begin{equation}
\tilde{\omega}_m^2 \approx \tilde{\omega}_m^{(0)2} \frac{1}{(1 - \epsilon \beta_{\omega }^{mm})}.\label{high_order_expansion}
\end{equation}

It can be shown~\cite{POM-2007}, that the high order perturbation expansion given in equation~(\ref{high_order_expansion}) remains valid even for $\epsilon \gg 1$. 

\begin{figure}

\centering
\includegraphics[width=14cm,height=4.cm]{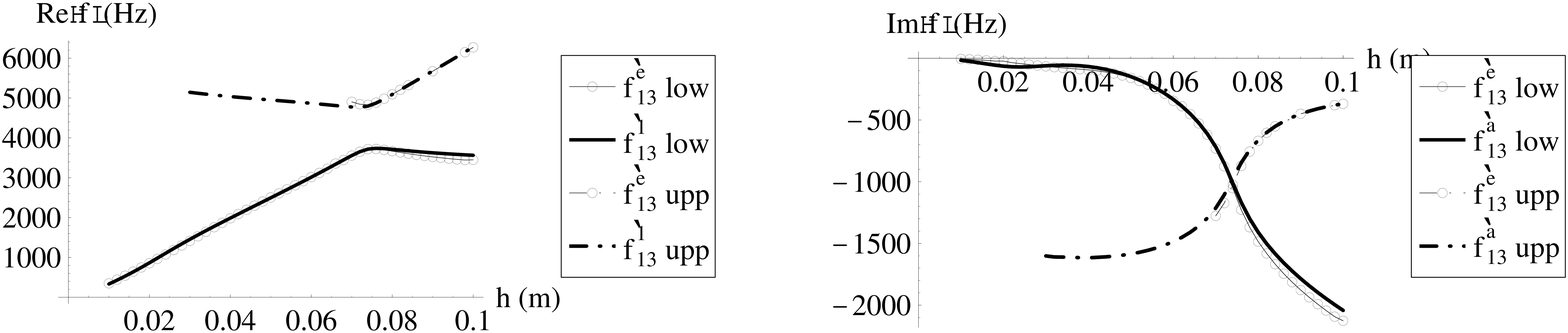} 
\caption{\footnotesize Comparison between the real (left) and imaginary (right) parts, both in Hertz, of the complex resonance frequency $f$ of the mode $(m=1,n=3)$ in the case of a steel plate in contact with water when varying its thickness $h$ in meter, obtained by using the exact solution (thin curves with circles) and the high order perturbation expansion (thick curves).} \label{fig_mode13}
\end{figure}

However, if one is interested in the resonance modes $\hat{U}_m(M)$ and resonances pulsations $\hat{\omega}_m$, given by $\hat{\Lambda}_m =\rho_p h \hat{\omega}_m^2$, that are the non-vanishing solutions of equation $a(\hat{U}_m,V^*) - \rho_p h \hat{\omega}_m^2 \left( \langle \hat{U}_m,V^* \rangle - \epsilon  \beta_{\hat{\omega}_m} (\hat{U}_m,V^*)  \right) =  0$ calculated using perturbation expansion~(\ref{high_order_expansion}), it can be easily seen that one has to solve the following resonance equation 
\begin{equation}
\hat{\omega}_m^2 = \tilde{\omega}_m^2 (\hat{\omega}_m ) \Leftrightarrow \hat{\omega}_m^2 \approx \tilde{\omega}_m^{(0)2}/\left(1 - \epsilon \beta_{\hat{\omega}_m }^{mm}\right) \Leftrightarrow \hat{\omega}_m^2 \left(1 - \epsilon \beta_{\hat{\omega}_m }^{mm}\right) \approx \tilde{\omega}_m^{(0)2}.
\end{equation}
This calculation is extremely difficult because one needs to find the complex roots of a non convex complex function. When attempting to solve this equation, the following cases occur:
\begin{itemize}
\item For $\epsilon = 0$, there is obviously one real resonance (and its opposite) solution of 
\begin{equation}
 \hat{\omega}_m^2 = \tilde{\omega}_m^2 = \tilde{\omega}_m^{(0)2}.
\end{equation}
\item For $\epsilon \ll 1$,  the fluid loaded resonance pulsations are very close to the in vacuo ones. And it can be shown that there there is one complex resonance (and the opposite of its conjugate $-\hat{\omega}_m^*$) solution of the approximate equation in which the modal impedance is estimated at the in vacuo eigenpulsation: 
\begin{equation}
 \hat{\omega}_m^2 = \tilde{\omega}_m^{(0)2} \left(1 + \epsilon \beta_{\tilde{\omega}_m^{(0)} }^{mm}\right),
\end{equation}
this relation is thereafter named the classical perturbation expansion resonance equation. 
\item For $\epsilon = O (1)$, there are one or more complex resonances (and the opposite of the corresponding conjugates), as shown in figure~(\ref{fig_mode13}) that are solutions of
\begin{displaymath}
 \hat{\omega}_m^2 = \tilde{\omega}_m^{(0)2}/\left(1 - \epsilon \beta_{\hat{\omega}_m }^{mm}\right),
\end{displaymath}
or, equivalently 
\begin{equation}
 \hat{\omega}_m^2 \left(1 - \epsilon \beta_{\hat{\omega}_m }^{mm}\right)-\tilde{\omega}_m^{(0)2} = 0, \label{high-order}
\end{equation}
this relation is thereafter named the high order perturbation expansion resonance equation.
\end{itemize}

The corresponding resonance modes are simply given by the calculating the eigenmodes at the corresponding resonance frequency $\hat{U}_m(M) = \tilde{U}_m (\hat{\omega}_m,M)$. 

As an example, in the case of a clamped steel plate measuring $1\mbox{m} \times 0,7\mbox{m}$ in water, the resonance frequency in the case of the mode $m=1$ $n=3$ calculated by solving the exact equations using a Boundary Element Method can be compared in figure \ref{fig_mode13} with those obtained using the high order perturbation method given by the numerical resolution of the high order perturbation expansion resonance equation~(\ref{high-order}) when varying the thickness of the plate. It is worth noting that the exact solution presented in figure \ref{fig_mode13} required approximately 1 month to calculate on a 4 processor parallel computer whereas, the calculation of the approximate solution required only a few minutes on the same computer using Mathematica~\cite{Mathematica}. In a vacuum, the real part of the frequency increases with thickness, while the imaginary part is zero; also as $\tilde{\omega}_m^{(0)2} \propto h^2$ and $\epsilon \propto 1/h$, the function $\omega^2 - \tilde{\omega}_m^{(0)2}/\left(1 - \epsilon \beta_{\omega}^{mm}\right) \approx \omega^2 $ for $h \rightarrow 0$. At a given thickness, the mode $m=1$ $n=3$ has at least two resonance frequencies located on either an ``upper branch'' (discontinuous curves) or a ``lower branch'' (continuous curves). It is worth noting that this behavior occurs at all the resonance modes of this plate. 

These results show not only that the behavior of this mode can be described very precisely using the high order perturbation expansion, but also (and this is the key point in the present paper) that a given resonance mode can have several resonance frequencies.

 
\section{MULTIPLE RESONANCES} \label{multiple}

Here we use the analytical expression for the resonances given by the high order perturbation expansion that is the various roots of equation~(\ref{high-order}), that is the complex pulsations $\omega=\hat{\omega}_m$ such that $\omega^2 (1 - \epsilon \beta_{\omega }^{m}) =  \tilde{\omega}_m^{(0)2}$. The whole problem focus here on the existence and the counting of these roots. 

The questions arising about the existence and counting of roots can be answered by using some classical results based on the theory of entire functions~\cite{Saks-Zygmund-1965, Levin-1996}. There is a strong link between the distribution of the zeros of entire functions and their order. Let us recall that if $f(z)$ is an entire function, the function $M(k) = \max_{|z| = k} |f(z)|$ will increase indefinitely together with $k$. The rate of growth of $M(k)$ can be characterized by comparing it with function $\exp(k)$. By definition, the order (of growth) $\rho$ is the smallest number such that $M(k) \le \exp k^{\rho+\delta}$ for every $\delta >0$ and for $k > k_0(\delta)$. From this, it can be easily seen that the order is given by the formula $\rho = {\lim \sup}_{k \rightarrow \infty} \ln \ln M(k) / \ln k$ ; as examples, a polynomial in $z$ is of order 0 and the exponential function is of order 1. Moreover, the Jensen formula (\cite{Saks-Zygmund-1965, Levin-1996}) states that if $n_{f} (k)$ is the number of zeros of an entire function $f(z)$ of order $\rho$ inside a disk with a radius $k$ centered at the origin, then if $k$ is sufficiently large, one has $n_f (k) \le  k^{\rho+\delta}$, $\delta>0$ (see~\cite{Saks-Zygmund-1965}, theorem (8.3) p.328). Since $n_f (k)$ is a non-decreasing function of $k$ it has either an infinite number of zeros or none (as in the case the function $\exp(z)$, which is never zero and is of order 1). 

Now the key point here is how to determine the order of the complex function $f(k) = k^2 - k_m^2 - \epsilon k^2 \beta^m(k) = 0 $, with $\beta^m(k) = \beta_{k }^{mm}$ and $k_m$ if a fixed real number. Since $k^2 - k_m^2$ and $k^2$ are zero order functions, using theorem (6.4)  in \cite{Saks-Zygmund-1965}: \textsl{if $F_1(z)$ and $F_2(z)$ are entire functions of order $\rho_1$ and $\rho_2$, respectively, and if $\rho_1 < \rho_2$ then the order $\rho$ of the sum $F_1(z) + F_2(z)$ is equal to $\rho_2$} and theorem (6.7) in \cite{Saks-Zygmund-1965}:\textsl{if $F(z)$ is an entire function of order $\rho$, and $P(z)$ is a polynomial of positive degree, then the product $F(z) P(z)$ has order $\rho$}, it is easy to see that the order of $f(k)$ depends only on the order of $\beta^m(k)$ when $\epsilon \neq 0$.

Now the next point is to determine the order of the radiation impedance $\beta^m(k)$ as an entire function of $k$. For this purpose, one shows with the simple example of a simply supported plate (this case was chosen because of the relative simplicity of the calculations involved) that each mode has an infinite number of resonance frequencies.

\subsection{The Simply Supported Plate}

Now let us take the more complex case of a simply supported rectangular plate (dimensions $a \times b$) vibrating in an $m \times n$ mode. In a vacuum, the mode is given by $\tilde{U}_{mn}^{(0)}(x,y) = 2/\sqrt{ab} \sin m \pi x /a \sin n \pi y /b$. The key point is again how to determine the order of the radiation impedance. The procedure developed here is slightly different from the previous one because is seems to be impossible to construct an entire radiation impedance series in terms of to the variable $k$ ; it is based on a method involving combinations of entire functions~\cite{Saks-Zygmund-1965}. By making a change of variable~\cite{Mangiarotty}, it can be easily shown that $\beta_{k }( \tilde{U}_{mn}^{(0)}, \tilde{U}_{mn}^{(0)*} ) = \beta^{mn}(k)$ is given by 
\begin{eqnarray}
\beta^{mn} (k) & \!\! = \!\! & - \frac{a b }{\pi} \int_0^1 \int_0^1 \int_0^1 \int_0^1 \sin m \pi x \sin n \pi y  \sin m \pi x' \sin n \pi y' \frac{e^{\imath k D}}{ \pi D} d x d y d x' d y'~~\\
               & \!\! = \!\! & - \frac{4 a b }{\pi} \int_0^1 \int_0^1 G_m (X) G_n (Y) \frac{e^{\imath k R}}{ \pi R} d X d Y,
\end{eqnarray}
where $D = \sqrt{a^2(x-x')^2+b^2(y-y')^2}$ and $R = \sqrt{a^2 X^2+b^2Y^2}$. The functions $G_m (x)$ and $G_n (y)$ are given by single analytic integrals 
\begin{eqnarray}
G_m (x) & = & \int_0^1 \sin m \pi (x+x') \sin m \pi x' d x' \\
        & = & \frac{ \sin m \pi x + m \pi(1-x) \cos m \pi x }{2 m \pi }
\end{eqnarray}
Now let us make a new change of variables from rectangular to polar coordinates
\begin{eqnarray}
\beta^{mn} (k) & = & - \frac{4 }{\pi} \int_0^a \int_0^b G_m \left( \frac{x}{a} \right) G_n \left( \frac{y}{b} \right) \frac{e^{\imath k r}}{ r} d x d y \\
               & = & - \frac{4 }{\pi} \int_0^{\theta_{\alpha}} J_{1mn}(k,\theta) d \theta  - \frac{4 }{\pi} \int_{\theta_{\alpha}}^{\pi/2} J_{21mn}(k,\theta) d \theta,
\end{eqnarray}
with 
\begin{eqnarray}
J_{1mn}(k,\theta) & = & \int_0^{\frac{a}{\cos \theta}} G_m \left( \frac{r \cos \theta}{a} \right) G_n \left( \frac{r \sin \theta}{b} \right)  e^{\imath k r} d r \\
J_{2mn}(k,\theta) & = & \int_0^{\frac{b}{\sin \theta}} G_m \left( \frac{r \cos \theta}{a} \right) G_n \left( \frac{r \sin \theta}{b} \right) e^{\imath k r} d r
\end{eqnarray}
The closed formulas obtained for both $J_{1mn}(k,\theta)$ and $J_{2mn}(k,\theta)$ using Mathematica~\cite{Mathematica} analytical integration contains thousands of terms (they are about 850 A4 pages each), making it impossible to be controlled or used. Then an approximation of these functions that preserves they order property of $\beta^{mn} (k)$ had been build. This approximation is denoted by $\check{\beta}^{mn} (k)$.

To construct this approximation, one can remark that the function to be studied is the sum of functions with of following form: \begin{equation}
F(k) \propto \int_0^{\theta_{\alpha}} \int_0^{g(\theta)} f(r,\theta) \exp(\imath k r) dr d\theta,
\end{equation}
The exact expression of $g(\theta)$ does not matter, what is important is that this function does not depend on $k$. Now by using integration by parts with respect to $r$ one obtains 
\begin{eqnarray}
F(k) & \propto & \int_0^{\theta_{\alpha}} f(g(\theta),\theta) \frac{e^{\imath k g(\theta)} }{\imath k} d \theta + {\cal O} (k^2) \\
     & \propto & \int_0^{\theta_{\alpha}} \frac{f(g(\theta),\theta)}{g^{\prime}(\theta)} \frac{e^{\imath k g(\theta)}}{\imath k} g^{\prime}(\theta) d \theta + {\cal O} (k^2),
\end{eqnarray}
again by using integration by parts with respect to $\theta$ one obtains:
\begin{equation}
F(k) \propto H(\theta_{\alpha}) \frac{e^{\imath k g(\theta_{\alpha})}}{k^2} + {\cal O} (k^3),
\end{equation}
with $H(\theta) =  f(g(\theta),\theta)/g^{\prime}(\theta)$. Then the order of the function $F(k)$, which is given by the rate of growth of the function $M(k) = \max_{|k|\rightarrow \infty} F(k)$, is given by $M(k) = \max_{|k|\rightarrow \infty} H(\theta_{\alpha}) \exp(\imath k g(\theta_{\alpha})) /k^2$, then the order does not depend on a exact description of the functions $H(\theta_{\alpha})$ and $g(\theta_{\alpha})$ and the approximations can be made on these functions without changing the impedance order.

Since one always has $0< \theta_{\alpha}< \pi/2$, one can expand the functions to be integrated with respect to the angular variable $\theta$ up to the order 1: $J_{1mn}(k,\theta)$ around $\theta=0$ and $J_{2mn}(k,\theta)$ around $\theta=\pi/2$. This gives
\begin{eqnarray}
J_{1mn}(k,\theta) \approx j_{1m}(k) & = & a \frac{ \imath a^3 k^3 m \pi + 2 m^3 \pi^3 - \imath a k m^3 \pi^3 - (-1)^m 2 \emath^{\imath a k} m^3 \pi^3 }{ 4 m \pi (a k - m \pi)^2 (a k + m \pi)^2 } \\
J_{2mn}(k,\theta) \approx j_{2n}(k) & = & b \frac{ \imath b^3 k^3 n \pi + 2 n^3 \pi^3 - \imath b k n^3 \pi^3 - (-1)^n 2 \emath^{\imath b k} n^3 \pi^3 }{ 4 n \pi (b k - n \pi)^2 (n k + n \pi)^2 }
\end{eqnarray}
$j_{1,2m}(k)$ are regular entire functions of $k$, in particular, it can easily be shown that
\begin{eqnarray}
j_{1m}(k=\pm m\pi/a) & = & \frac{ 3 \imath a^2 m^2 \pi^2 + a^2 m^3 \pi^3 }{ 16 a^2 m^3 \pi^2 } \\
         j_{1m}(k=0) & = & a \frac{ 2 m^3 \pi^3 -2 (-1)^m m^3 \pi^3 }{4 m^5 \pi^5 } \\
j_{2n}(k=\pm n\pi/b) & = & \frac{ 3 \imath b^2 b^2 \pi^2 + b^2 n^3 \pi^3 }{ 16 b^2 n^3 \pi^2 } \\
        j_{2n}(k=0) & = & b \frac{ 2 n^3 \pi^3 -2 (-1)^n n^3 \pi^3 }{4 n^5 \pi^5 }
\end{eqnarray}

Then, by integrating with respect to $\theta$, one obtains $\check{\beta}^{mn} (k) = -4/\pi ( j_{1m}(k) \theta_{\alpha} + j_{2n}(k)(\pi/2-\theta_{\alpha}))$. It is worth noting that the accuracy of the approximation of $\beta^{mn} (k)$ decreases as the mode order increases. However, at low orders (say $m \le 3$ $n \le 3$) the accuracy is very good.  Now as $j_{1m}(k)$ and $j_{2n}(k)$ are entire functions of $k$, $\check{\beta}^{mn} (k)$ is also an entire function of $k$. It still remains to estimate the order of $\check{\beta}^{mn} (k)$. For given $m$ and $n$ and for $|k| \rightarrow \infty$, one has
\begin{eqnarray}
j_{1m}(k) & \approx & \frac{ - (-1)^m 2 m^2 \pi^2 \emath^{\imath a k}  }{ a^3 k^4 } \\
j_{2n}(k) & \approx & \frac{ - (-1)^n 2 n^2 \pi^2 \emath^{\imath b k}  }{ b^3 k^4 },
\end{eqnarray}
and  
\begin{equation}
\check{\beta}^{mn} (k) \approx \frac{2}{\pi^2} \left( (-1)^m  \frac{m^2}{a^2} \theta_{\alpha} \emath^{\imath a k} + (-1)^n \frac{n^2}{b^2} \left(\frac{\pi}{2}- \theta_{\alpha} \right) \emath^{\imath b k} \right) \frac{ 1  }{ k^4 }.
\end{equation}
The order of $\check{\beta}^{mn} (k)$ is easy to obtain from classical results on the sum and product of entire functions (see for example \cite{Saks-Zygmund-1965}, chapter VII paragraph 6). First it is worth noting that the entire function $(-1)^m  m^2/a^2 \theta_{\alpha} \emath^{\imath a k} + (-1)^n n^2/b^2 \left(\pi/2- \theta_{\alpha} \right) \emath^{\imath b k}$ cannot be identically zero for any value of $a$, $b$, $m$ or $n$ when $|k| \rightarrow \infty$. To see this more clearly, one can deal separately with the cases $a=b$ and a$\neq b$.

\paragraph{The case $a = b$.}
One has $\theta_{\alpha} = \pi/4$ and $\check{\beta}^{mn} (k) \approx \frac{1}{2 \pi a^2} \left( (-1)^m m^2 + (-1)^n n^2 \right) \emath^{\imath a k} \frac{ 1 }{ k^4 }$ becomes the product of the exponential function $\emath^{\imath a k}$ of order 1 and the inverse of a degree 4 polynomial in $k$, which is of order 0. $\check{\beta}^{mn} (k)$ is therefore an entire function of order 1. 

\paragraph{The case $a \neq b$.}
Let us take for example $a>b$. Then with $|k| \rightarrow \infty$ one has $\max \left|\emath^{\imath a k}\right| \gg \max \left|\emath^{\imath b k}\right|$; it is  worth noting that this is true in the half plane $\Im (k) <0$, whereas in the half plane $\Im (k) > 0$, $\check{\beta}^{mn} (k)$ tend rapidly to zero. In the half plane $\Im (k) <0$, one therefore has $\check{\beta}^{mn} (k) \approx \frac{2}{\pi^2} \left( (-1)^m  \frac{m^2}{a^2} \theta_{\alpha} \emath^{\imath a k} \right) \frac{ 1  }{ k^4 }$ which is also the product of the exponential function $\emath^{\imath a k}$ of order 1 and the inverse of a degree 4 polynomial in $k$, which is of order 0. $\check{\beta}^{mn} (k)$ is therefore an entire function of order 1.

From the Jensen formula, the number of zeros in an entire function $f(z)$ or order 1 inside the disk with radius $r$ centered at the origin, if $r$ is sufficiently large, is given by $n(k) \le k^{1+\delta}$, $\delta>0$ and since $n (k)$ is a non-decreasing function of $k$ it has either an infinite number of zeros or none. Then, with $|k| \rightarrow \infty$, the radiation impedance of the membrane has either an infinite number of zeros or none. To establish that there is at least one zero (and then an infinite number), one applies the argument principle whereby if $C$ is a disk  with radius $K$ centered on the origin, the number of zeros inside this disk $n_{\check{\beta}}(K)$ is given by the integral $n_{\check{\beta}}(K) = 1/(2 \imath \pi) \oint_{C} \check{\beta}^{\prime} (k)/\check{\beta}(k) d k, \mbox{ with } k = K e^{\imath \theta}$. 

The number of zeros inside this disk, in the case of the mode $m n$, denoted by $n_{\check{\beta}^{mn}}(K)$, is given by the integral $n(K) = 1/(2 \imath \pi) \oint_{C} \check{\beta}^{\prime mn} (k)/ \check{\beta}^{mn} (k) d k, \mbox{ with } k = K e^{\imath \theta}$, if $\check{\beta}^{mn} (k)$ has no pole inside $C$. For example, with $a= 1$ and $b=0.7$, one computes  the previous integral numerically. It can easily be established that in the mode $m=1$, $n=1$ $n_{\check{\beta}^{11}} (11) = 0$, $n_{\check{\beta}^{11}} (12) = 2$, $n_{\check{\beta}^{11}} (18) = 4$ and in the mode $m=1$, $n=3$ $n_{\check{\beta}^{13}}(7) = 0$, $n_{\check{\beta}^{13}}(8) = 2$, $n_{\check{\beta}^{13}}(19) = 4$. Again, the modal impedance has an infinite number of zeros. Therefore, while in a vacuum, each mode of a simply supported plate has only one resonance frequency (and its opposite), under heavy loading condition each mode will have an infinite number of resonance frequencies. 

\begin{figure}
\centering
\includegraphics[width=15.5cm,height=6cm]{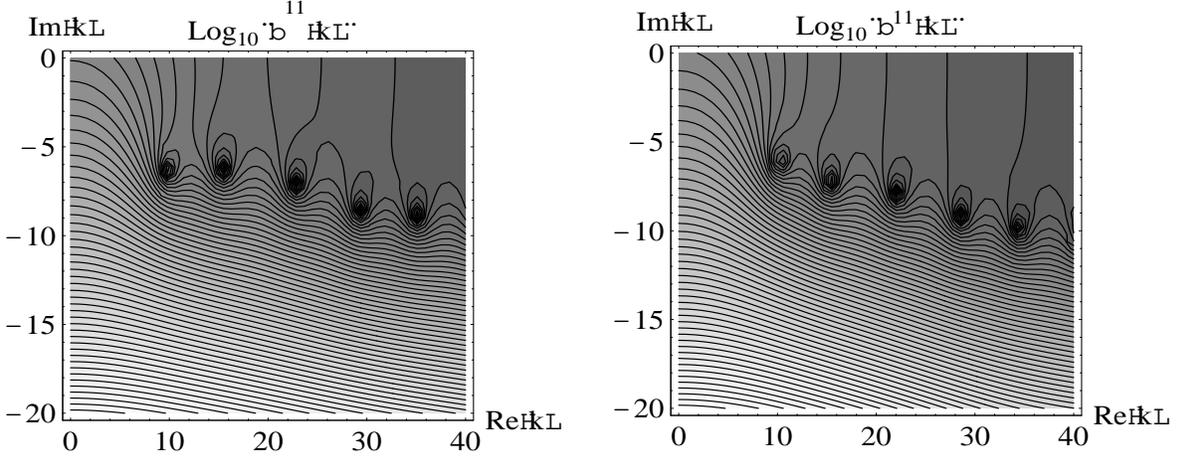}
\caption{\footnotesize Contour plots of the amplitude of the radiation impedance in the mode $m=1$, $n=1$ in the case of a simply supported plate ($1 \mbox{ m } \times 0.7 \mbox{ m }$). Left: approximate $\check{\beta}^{11}( k)$, right: exact $\beta^{11} ( k)$. } \label{fig_beta_mode11}
\end{figure}

\begin{figure}
\centering
\includegraphics[width=15.5cm,height=6cm]{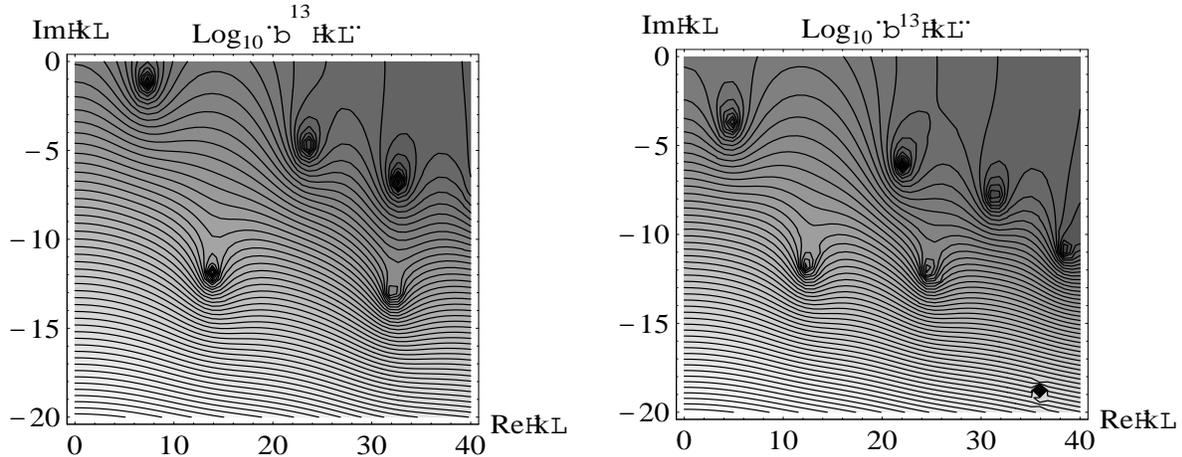}
\caption{\footnotesize Contour plots of the amplitude of the radiation impedance in the mode $m=1$, $n=3$ modes in the case of a simply supported plate ($1 \mbox{ m } \times 0.7 \mbox{ m }$). Left: approximate $\check{\beta}^{13}( k)$, right: exact $\beta^{13} ( k)$. } \label{fig_beta_mode13}
\end{figure}

Figures~\ref{fig_beta_mode11} and~\ref{fig_beta_mode13} give contour plots for the amplitude of the radiation impedance $\beta^{mn}( k)$ and its approximation $\check{\beta}^{mn} (k)$ in the mode $m=1$ $n=1$ in figure~\ref{fig_beta_mode11} and in the mode $m=1$ $n=3$ in figure~\ref{fig_beta_mode13}. As previously, on these curves, the dark spots correspond to the zeros of the radiation impedances. It is worth noting that the lower the mode order, the better the approximation $\check{\beta}^{mn} (k)$ becomes.

It is worth noting that all the figures~\ref{fig_beta_mode11} to \ref{contour_high_order} are plotted in a quarter of the complex plane. The upper half complex plane has no root: the frequency dependence chosen $\exp( -\imath \omega t)$ implies a complex pulsation $\omega$ with negative imaginary part to ensure a decrease of the solution for $t \rightarrow \infty$. The curves are plotted on the quarter lower plane with either positive or negative real part without any particular difference since the various resonances are symmetric with respect to the imaginary axis: resonances occur alway by conjugate pairs of the form $\hat{\omega}_m$ and $-\hat{\omega}_m^*$ when $\hat{\omega}_m \in \Cmath$ or $\hat{\omega}_m$ and $-\hat{\omega}_m$ when $\hat{\omega}_m \in \Rmath$

\subsection{Comments About the General Case}

Now let us return to the initial question of the existence or non-existence of multiple resonances in fluid-loaded plates. Let us recall that in the case of heavy loading, one has to look for the resonance pulsations $\hat{\omega}_{mn}$ such that $\hat{\omega}_{mn}^2 (1 - \epsilon \beta_{\hat{\omega}_{mn} }^{mn}(\omega)) - \tilde{\omega}_{mn}^{(0)2} = 0$. As shown in the previous paragraphs, this behavior occurs in a large class of problems. Although this statement has not been proved to be be true under all possible boundary conditions, it seems reasonable to assume that under boundary conditions of all kinds and under very heavy loading condition every resonance mode in a plate has an infinite number of resonance frequencies.

\begin{figure}
\centering
\includegraphics[width=15.5cm,height=6cm]{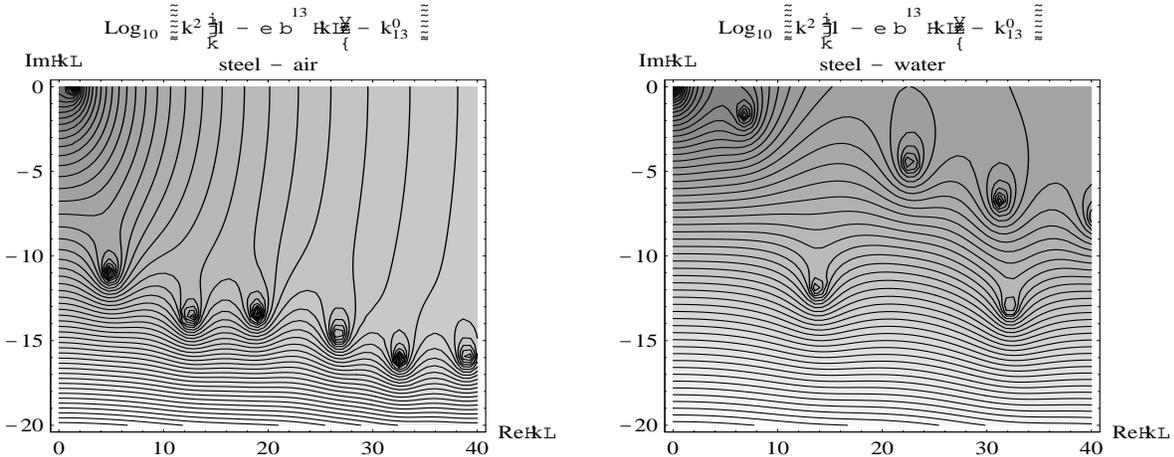}
\caption{\footnotesize Contour plots of the amplitude of the resonance value equation $k^2 (1 - \epsilon \beta^{mn}(k)) -  k_{mn}^{(0)2}$ in the mode $m=1$, $n=3$ in the case of a thin simply supported plate ($1 \mbox{ m } \times 0.7 \mbox{ m } \times 2 \mbox{ mm}$). Left: steel plate in contact with air ($\epsilon = 0.08$), right: steel plate in contact with water ($\epsilon = 64$). } \label{fig_eigenvalue_2mm_13_air_water}
\end{figure}

But in the case of light fluid loading conditions, as shown with the previous results whatever the loading conditions, the resonance value relation has several zeros. One close to the real axis with a small imaginary part a a set of resonance with very large imaginary part (see left curve in figure~\ref{fig_eigenvalue_2mm_13_air_water}). The later are very difficult to detect because of their large imaginary part tha implies a very rapid vanishing of their contribution.  But as shown by Sanchez~\cite{Sanchez-Sanchez-1989}, under light loading condition (as in the case of a steel plate in contact with air, for example), there exist only the resonances of the structure alone (one in each mode shifted toward the complex plane) and the complex scattering resonances of the fluid in contact with the rigid solid. This is an apparent contradictory situation because Sanchez deals with a slightly different problem since he looks for scattering frequencies for a rigid structure in vacuum with zero eigenvalue (that is $\tilde{\omega}_{mn}^{(0)2} = 0$) which has only a set of scattering resonance frequencies.

The results presented in left curve in figure~\ref{fig_eigenvalue_2mm_13_air_water} also show that except for the zero close to origin, the zeros of the resonance value relation corresponding to the plate loaded with water are very similar to those given by the radiation impedance. In that case, contrarily to the light fluid loading situation, both structure resonance and scattering resonances have their imaginary part with comparable magnitude, that allows a possible identification of the corresponding modes.

Now under moderate loading condition (as in the case of a steel plate in contact with water), what is the relevant relation ? This seems to be  an open question. Although the resonance relation $k^2 (1 - \epsilon \beta^{mn}(k)) -  k_{mn}^{(0)2} = 0$ is true under very heavy loading, it has to be supplemented under small or moderate loadings conditions (that is, when $\epsilon = {\cal O} (1)$) by terms accounting for the cross modal impedance, which make it almost impossible to conduct the above analysis.

\subsection{Numerical Example on a Clamped Steel Plate in Water}

\begin{figure}
\centering
\includegraphics[width=10cm,height=8cm]{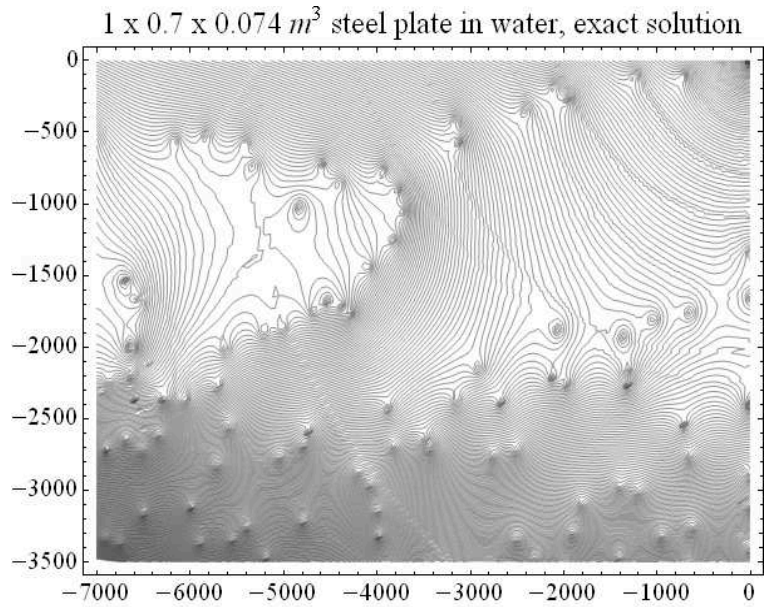}
\caption{\footnotesize Contour plots of the Logarithm of the determinant amplitude of the linear system of simultaneous equation obtained by solving the Boundary Integral Equation Method for a clamped steel plate in water. } \label{contour_exact}
\end{figure}

\begin{figure}
\centering
\includegraphics[width=10cm,height=8cm]{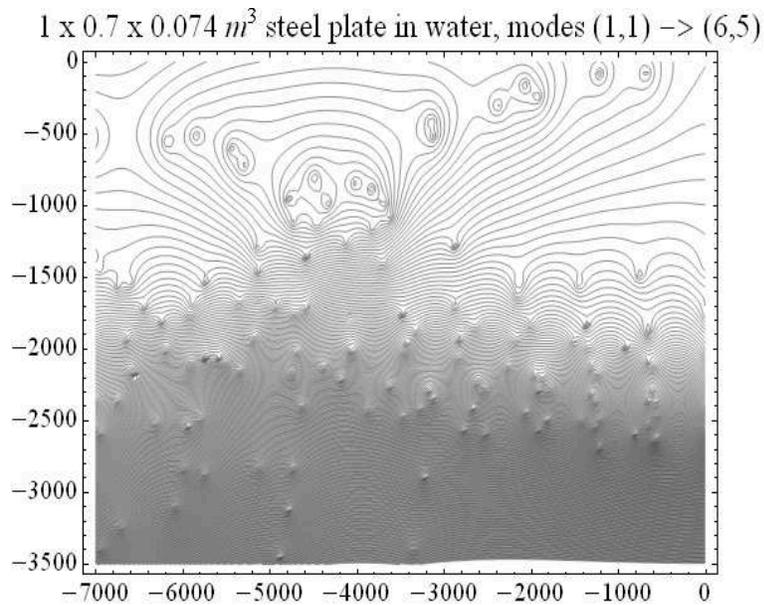}
\caption{\footnotesize Contour plots of the Logarithm of the product of the thirty firsts high order resonances equations (corresponding to the modes $(1,1)$ to $(6,5)$) for a clamped steel plate in water. } \label{contour_high_order}
\end{figure}

To conclude this analysis, one presents in figure~\ref{contour_exact} and figure~\ref{contour_high_order} a comparison of the location of the resonances for a clamped steel plate in contact with water, using contour plots. In these curves the various dark spots show a rough estimate of the roots. In figure~\ref{contour_exact} it is showed a contour plot of the Logarithm of the determinant amplitude of the linear system of simultaneous equation obtained by solving the exact Boundary Integral Equation Method using a Tchebycheff-collocation method~\cite{POM-1996} and in figure~\ref{contour_high_order}, it is plotted the contour plot of $F(\omega)$ defined as a product of the thirty firsts high order resonances equations (corresponding to the modes $(1,1)$ to $(6,5)$) for the same plate: $F(\omega)=\prod_{m= 1}^{n=1} \prod_{m= 6}^{n=5} \omega^2 \left(1 - \epsilon \beta_{\omega }^{{mn}}\right)-\tilde{\omega}_{mn}^{(0)2}$ which roots are obviously those of the corresponding high order resonances equations. While the absolute values (and contour lines) can not be directly compared, the approximate location of the roots of the function $F(\omega)$ gives a reasonable image of the true resonances. These figures show also that the spectrum of this fluid-loaded structure is not a combination of the resonances of the structure alone (with frequency lowered by added mass and shifted through the complex plane by radiation damping) and of the scattering resonances of the rigid body alone. Each scattering resonances set is shifted through the complex plane differently by each resonance mode of the structure allowing a very complex distribution of the resonances of the fluid-loaded structure. It is worth noting that this distribution of the resonances makes a direct numerical estimate of the spectrum quite impossible. This also shows that there is always the need of analytical estimates such as high order perturbation expansion to be able to construct a numerically efficient estimate of the true spectrum.

\section{CONCLUSIONS} \label{conclusions}
Here we have shown that the loading of a rectangular plate with a high density fluid transforms each resonance into an infinite number of resonances. These results open up new prospects. In particular, it would be interesting to obtain similar estimates on the vibratory modes occurring in clamped plates or mechanical structures of other kinds, although these may not be easy to obtain. Another interesting aspect concerns the asymptotic properties of the zeros of entire functions. As we have seen  above, making spectral estimates is equivalent to finding the roots of the function $f(k) = k^2 (1 - \epsilon \beta^{mn}(k)) -  k_{mn}^{(0)2}$, but $f(k)$ is an order 1 function at $\epsilon \neq 0$, and it can easily be shown that it is also a function of normal type~\cite{Levin-1996}, then by definition $f (k)$ is an entire function of exponential type. Entire functions of this kind which satisfy the inequality $\int_{-\infty}^{+\infty} \ln^+ |f (k)| /(1+k^2) d k < \infty$ belongs to class ${\cal A}$~\cite{Levin-1978} and have regularity properties as regards the asymptotic distribution of their zeros. Another question which arises is how to determine under what conditions moderately coupled systems (such as a steel plate in contact with water) will have these properties. Another important aspect worth investigating is the validity of modal expansions in the case of resonance modes having two or more resonances frequencies. It would also be interesting to develop experimental device which would make it possible to test the occurrence of the behavior described here.


\section{ACKNOWLEDGEMENTS}

This work was partially supported by the Agence Nationale de la Recherche (ANR) thru grant ANR-06-BLAN-0081-01.


\end{document}